%% file: comb_prl.tex
\documentclass[aps,prl,showpacs,twocolumn,lineno,groupedaddress]{revtex4}

\usepackage{graphicx}
\usepackage{epsfig}
\usepackage{dcolumn}
\usepackage{bm}
\usepackage{amssymb}   
\usepackage[usenames]{color} 

\newcommand{\beq}{\begin{equation}}
\newcommand{\eeq}{\end{equation}}
\newcommand{\barr}{\begin{eqnarray}}
\newcommand{\earr}{\end{eqnarray}}

\def\bq{\begin{quote}}
\def\eq{\end{quote}}

\def\spose#1{\hbox to 0pt{#1\hss}}
\def\lsim{\mathrel{\spose{\lower 3pt\hbox{$\mathchar"218$}}
 \raise 2.0pt\hbox{$\mathchar"13C$}}}
\def\gsim{\mathrel{\spose{\lower 3pt\hbox{$\mathchar"218$}}
 \raise 2.0pt\hbox{$\mathchar"13E$}}}

\def\bsdec{${B_s^0 \rightarrow J/\psi \phi}$}

\def\D0{D\O }

\def\GeVp{ {\ifmmode \;{{\mbox{\mathrm GeV}} / {\mbox\mathrm c}} \else
${{\mbox{\mathrm GeV}} / {\mbox\mathrm c}}$ \fi }}
\def\MeVp{ {\ifmmode \;{{\mbox{\mathrm MeV}} / {\mbox\mathrm c}} \else
${{\mbox{\mathrm MeV}} / {\mbox\mathrm c}}$ \fi }}
\def\MeV{ {\ifmmode \;{{\mbox{\mathrm MeV}} / {\mbox\mathrm c}^2} \else
${{\mbox{\mathrm MeV}} / {\mbox\mathrm c}^2}$ \fi }}
\def\GeV{ {\ifmmode \;{{\mbox{\mathrm GeV}} / {\mbox\mathrm c}^2} \else
${{\mbox{\mathrm GeV}} / {\mbox\mathrm c}^2}$ \fi }}

\begin{document}




\hspace{5.2in}
\mbox{FERMILAB-PUB-07-044-E}

\title{Combined D0 Measurements Constraining the CP-violating Phase and
Width Difference in the $B_s^0$ System  }
\date{February 20, 2007}

\input list_of_authors_r2.tex





           
\begin{abstract}

We combine the D0 measurement of 
the width difference between the light  and  
heavy  $B_s^0$ mass eigenstates
and of the CP-violating mixing phase determined from 
 the time-dependent angular
distributions in the \bsdec\ decays along with the charge asymmetry in 
semileptonic  decays also measured with the D0 detector. 
With the additional constraint from the
world average of the flavor-specific $B_s^0$  lifetime, we obtain  
$\Delta \Gamma_s \equiv (\Gamma_L - \Gamma_H)$
=  0.13 $\pm$ 0.09  ps$^{-1}$ and   
$\phi _{s} = -0.70 ^{+0.47} _{-0.39}$.
The data sample corresponds to an integrated luminosity of 1.1 fb$^{-1}$
accumulated with the D0 detector at the Fermilab Tevatron Collider.


\end{abstract}


\maketitle

\newpage

One of the great challenges  for  elementary particle physics
is to trace all possible sources of the violation of  
CP symmetry.
In the standard model (SM) of  particle physics, 
CP symmetry is violated through the CKM 
mechanism~\cite{ckm}.
Although the SM picture of CP violation has so far been confirmed by
all laboratory measurements, it has an unsolved problem: the level
of CP violation in the SM is too small to produce the observed baryon number
density in the universe~\cite{baryon}. Sources of CP violation beyond
the CKM mechanism must, therefore, exist to account for the
deficit. 
One source of  CP violation arises in the mixing of
doublets of neutral mesons. CP violation in $K^0$ mesons, composed of a
down quark and a strange quark, was discovered 40 years ago~\cite{K0}. 
The  $B_s^0$ mesons are similar 
quark-antiquark bound states, with the down quark replaced by 
a bottom quark.

In the SM, the light ($L$) and heavy ($H$) mass eigenstates 
of the mixed $B_s^0$ system  are expected
to have  sizeable 
mass and decay width differences: $\Delta M_s \equiv M_H - M_L$ and
$\Delta \Gamma_s \equiv \Gamma_L - \Gamma_H$. 
The two mass eigenstates are expected to be almost pure CP 
eigenstates.  The CP-violating mixing phase
is predicted~\cite{LN2006}
to be $\phi_s = (4.2 \pm 1.4)\times 10^{-3}$. 
New phenomena may alter  $\phi_s$
leading to a reduction of the observed $\Delta \Gamma_s $
compared to the SM prediction~\cite{LN2006}  $\Delta \Gamma_s^{SM} $:
$\Delta \Gamma_s$ =  $\Delta \Gamma_s^{SM} \times |\cos \phi_s|$.
While  $B_s^0$ --  $\bar B_s^0$ oscillations have
been detected   ~\cite{dmsd0} and
the mass difference
has recently been measured to  high precision~\cite{dms},
the CP-violating  phase remains unknown.
 The D0 experiment~\cite{run2det}  at the Fermilab Tevatron Collider
has 
conducted  a series of 
studies~\cite {psiphi,tausl,asldimu,aslbs} 
of $B_s^0$ mesons produced in proton-antiproton ($p\bar p$)
interactions.
This Letter utilises these results to obtain the best
estimate of the CP-violating  phase in the  $B_s^0$ system.

In Ref.~\cite{psiphi}, we studied the decay sequence \bsdec, 
$J/\psi \rightarrow \mu^+ \mu^-$, $\phi \rightarrow K^+K^-$.
From a fit to the time-dependent  angular distribution of 
the decay products,
we obtained  the mean lifetime, 
$\overline \tau_s =1/\overline {\Gamma}_s$  (where
$\overline{\Gamma}_s\equiv(\Gamma_H+\Gamma_L)/2$),
$\Delta {\Gamma}_s$, and 
the first direct constraint 
on $\phi_s$.
As discussed in Ref.~\cite{psiphi},
there is a 4-fold ambiguity in the result for $\phi_s$:
$\pm \phi_s$  and $\pm (\pi - \phi_s)$.
The sign of  $\sin\phi_s$ is reversed with the simultaneous reversal
of the signs of the cosines of the CP-conserving strong phases
 $\delta_1$ and  $\delta_2$. (We adopted the amplitude definition
and sign convention of Ref.~\cite{ddf}).
The possible  solutions are
\begin{eqnarray}
\label{solutions}
|\phi_s | & = & 
0.79 \pm 0.56 \rm~{(stat)} ^{+0.01}_{-0.14} \rm~{(syst)}, \nonumber \\
\Delta \Gamma_s & = &
 0.17 \pm 0.08 \rm~{(stat)} \pm 0.02 \rm~{(syst)} \rm~{ps}^{-1};\nonumber \\ 
\nonumber \\
|\phi_s| & = & 
2.35 \pm 0.56 \rm~{(stat)} ^{+0.14}_{-0.01} \rm~{(syst)}, \nonumber \\
\Delta \Gamma_s &=&
 -0.17 \pm 0.08 \rm~{(stat)} \pm 0.02 \rm~{(syst)} \rm~{ps}^{-1}; \nonumber \\ 
\nonumber \\
\
\end{eqnarray}
The first two solutions are consistent with the SM 
prediction~\cite{LN2006}.

\begin{figure}[h!tb]
\begin{center}\includegraphics[%
  width=7.8cm,
  keepaspectratio,
  trim=0 40 0 0
  ]{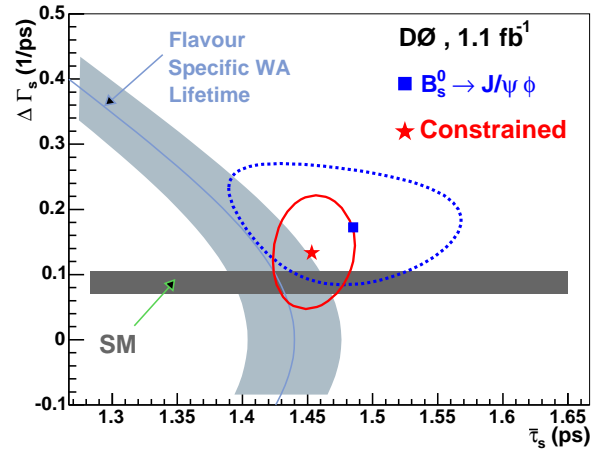}\end{center}
\caption{
The  error ellipse ($\Delta\ln({\cal L}) = 0.5$) 
in the plane $\Delta \Gamma_s$ versus $\overline {\tau}_s$
 for the
fit to the \bsdec\ data (dashed blue line) and
for the fit with the constraint from the two  D0 measurements
of the charge asymmetry in semileptonic $B_s^0$ decay,
and from the world average  flavor-specific lifetime
(solid red line). 
Also shown is a 
one-$\sigma$ 
band representing the world average result~\cite{hfag} for 
$\tau_{\text {fs}}$ and
a one-$\sigma$ band representing
the theoretical prediction
$\Delta \Gamma_s^{SM} = 0.088 \pm 0.017$ ps$^{-1}$~\cite{LN2006}.
[color online]
}
\label{fig:dgvstau}
\end{figure}

Flavor-specific decays of  $B_s^0$ mesons provide independent constraints
on the parameters of the system.
An effective  mean lifetime, resulting from a  single-exponential fit
to the decay time distribution,
$\tau_{\text {fs}}=1/\Gamma_{\text {fs}}$, is
 related to the physics parameters
$\overline \Gamma_s$ and $\Delta \Gamma_s$
 through the equation
$\Gamma_{\text {fs}}=\overline \Gamma_s - (\Delta\Gamma_s)^2/2\overline \Gamma_s + {\cal O}{(\Delta
 \Gamma_s)^3/\overline \Gamma_s^2} \label{gammafs}$~\cite{TeVbook}.
We use the world-average value,
$\tau_{\text {fs}}=1/\Gamma_{\text {fs}} = 1.440 \pm 0.036$ ps~\cite{hfag},
from a fit including 
the recent D0  measurement,
$\tau_{\text {fs}}=1/\Gamma_{\text {fs}} = 1.398 \pm 0.044$ (stat) $^{+0.028}_{-0.025}$ 
(syst) ps~\cite{tausl}.

Independently, we obtained another constraint on the parameters of the 
$B_s^0$ system from the measurements of the semileptonic charge
asymmetry induced by  $B_s^0$ mixing.
In general, the semileptonic charge asymmetry is defined as \cite{PDG}:
\begin{equation}
\label{asl}
A_{SL}^q = \frac{ N(\bar B_q^0 \to \ell^+ X) - N(B_q^0 \to \ell^- X) }
              { N(\bar B_q^0 \to \ell^+ X) + N(B_q^0 \to \ell^- X) }.
\end{equation}
It is related to the CP phase $\phi_q$ by \cite{nierste-b}:
\begin{equation}
\label{aslphi}
A_{SL}^q = \frac{\Delta \Gamma_q} {\Delta M_q} \tan{\phi_q}.
\end{equation}

In Ref.~\cite{asldimu}, we measured the same-sign dimuon charge asymmetry
defined as:
\begin{equation}
A_{SL}^{\mu \mu}  =  \frac{N(b \bar b \to \mu^+ \mu^+ X) - N(b \bar b \to 
\mu^- \mu^- X)}
         {N(b \bar b \to \mu^+ \mu^+ X) + N(b \bar b \to \mu^- \mu^- X)}.
\end{equation}
Both $B_d^0$ and $B_s^0$ contribute to this quantity~\cite{grossman},
and the result
of Ref.~\cite{asldimu} is given as:
\begin{eqnarray}
\label{asla}
A_{SL}^d & + & \frac{f_s Z_s}{f_d Z_d} A_{SL}^s =  \nonumber \\
& = & -0.0092 \pm 0.0044~\mbox{(stat)} \pm 0.0032~\mbox{(syst)}; \\
Z_q & = & \frac{1}{1-y_q^2} - \frac{1}{1+x_q^2}; \nonumber \\
x_q & = & \Delta M_q / \Gamma_q; ~~~~ y_q = \Delta \Gamma_q / (2 \Gamma_q);
\nonumber 
\end{eqnarray}
where $A_{SL}^d$ and  $A_{SL}^s$ are the charge asymmetries of the
$B_d^0$ and $B_s^0$ semileptonic decays, and
$f_d$ and $f_s$ 
are the production rates of  $B_d^0$ and $B_s^0$ mesons
in the hadronization of the $b$ quark, respectively.
In deriving  relation (\ref{asla}),  it is assumed
that there is no direct CP violation in semileptonic $B$ decays and that
the semileptonic width of all $B$ mesons is the same.
Using the world-average values~\cite{PDG}
$f_d = 0.398 \pm 0.012$, $f_s = 0.103 \pm 0.014$, 
$x_d = 0.776 \pm 0.008$, and
$Z_d = 0.376 \pm 0.006$  we obtain:
\begin{equation}
\label{coef}
\frac{f_s Z_s}{f_d Z_d} = 0.70 \pm 0.07~\mbox{(syst)} \pm 0.10~\mbox{(PDG)}.
\end{equation}
The value of $Z_s$ was computed using the measured values of 
$\Delta \Gamma_s$~\cite{psiphi},
$\overline {\tau}_s$~\cite{psiphi},
 and $\Delta M_s$~\cite{dms}: $Z_s = 1.015^{+0.018}_{-0.010}$. 
We have tested that propagating the  $Z_s$
dependence on $\Delta \Gamma_s$ has negligible effect on the
final results.
The systematic uncertainty arises mainly from a conservative
estimate of a possible variation in the reconstruction efficiency
of muons from semileptonic decays of different $B$ mesons~\cite{blife}.


The asymmetry $A_{SL}^d$ has been measured at $B$ factories 
where only $B_d^0$ and $B^\pm$ mesons are produced.
The average value of $A_{SL}^d$ is~\cite{hfag}: 
$A_{SL}^d = -0.0047 \pm 0.0046$. Combining this value and
(\ref{asla},\ref{coef}), and adding statistical and systematic uncertainties
in quadrature,
we obtain:
\begin{equation}
\label{aslc}
A_{SL}^s = -0.0064 \pm 0.0101.
\end{equation}

\begin{figure}[h!tb]
\begin{center}\includegraphics[%
  width=7.8cm,
  keepaspectratio,
  trim=0 40 0 0
  ]{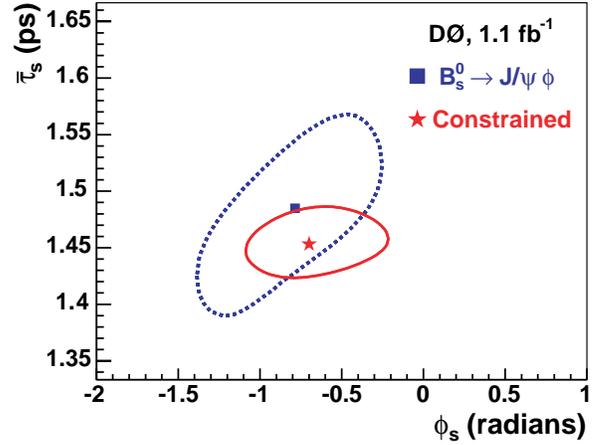}\end{center}
\caption{
The error ellipse ($\Delta\ln({\cal L}) = 0.5$) 
in the plane  ($\overline \tau_s $, $\phi_s$) for the
fit to the \bsdec\ data (dashed blue line) and
for the fit with the constraint from the two  D0 measurements
of the charge asymmetry in semileptonic $B_s^0$ decay,
and from the world average flavor-specific lifetime
(solid red line).[color online] 
}
\label{fig:bs_tauvsphi}
\end{figure}

\begin{figure}[h!tb]
\begin{center}\includegraphics[%
  width=7.8cm,
  keepaspectratio,
  trim=0 40 0 0
  ]{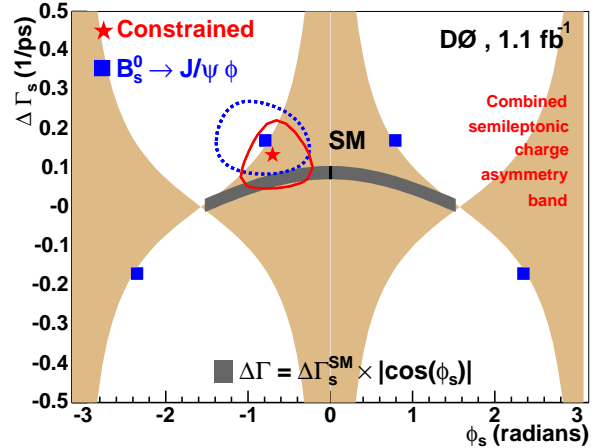}\end{center}
\caption{
The  error ellipse ($\Delta\ln({\cal L}) = 0.5$) 
in the plane  ($\Delta \Gamma_s $, $\phi_s$) for the
fit to the \bsdec\ data (dashed blue line) and
for the fit with the constraint from the two  D0 measurements
of the charge asymmetry in semileptonic $B_s^0$ decay,
and from the world average  flavor-specific lifetime
(solid red line). 
The central values for all four 
 solutions of the unconstrained fit are  indicated by  blue squares.
Also shown is the 
band representing the relation
$\Delta \Gamma_s$ =  
$\Delta \Gamma_s^{SM} \times |\cos \phi_s|$, with  
$\Delta \Gamma_s^{SM} = 0.088 \pm 0.017$ ps$^{-1}$~\cite{LN2006} 
(dark shade),
and the area corresponding to Eq.~\ref{aslg} (light shade) .
[color online]
}
\label{fig:bs_contdgvsdphi_aslcombtausl}
\end{figure}

In Ref.~\cite{aslbs}, we measured  $A_{SL}^s$ directly by
 using  all events with at least one muon 
that were consistent with the sequential decay
 $B_s^0 \to \mu \nu D_s$ with $D_s \to \phi \pi$.
The result of this  measurement is:
\begin{equation}
\label{asld}
A_{SL}^s = +0.0245 \pm 0.0193~\mbox{(stat)} \pm 0.0035~\mbox{(syst)}.
\end{equation}
The measurements (\ref{aslc}) and (\ref{asld}) are nearly independent
 since the fraction of dimuon final states in the sample of semileptonic
$B_s^0$ decays used in Ref.~\cite{aslbs} is only about 10\% \cite{bflavor}, 
and the fraction of 
semileptonic decays $B_s^0 \to \mu \nu D_s$ with $D_s \to \phi \pi$
in the dimuon sample used in Ref.~\cite{asldimu} is less than 1\%. 
Also, the systematic uncertainties of the two measurements
are uncorrelated. The main source of systematic
uncertainty in (\ref{aslc}) is the correction due to
$K^{\pm}$ decays, while in the case of the measurement (\ref{asld})
it is the fitting procedure.

Their combination gives the
best estimate of the charge asymmetry in semileptonic $B_s^0$ decays:
\begin{equation}
\label{asle}
A_{SL}^s = 0.0001 \pm 0.0090.
\end{equation}
Using relation (\ref{aslphi}) and the  result
$\Delta M_s = 17.8 \pm 0.1$ ps$^{-1}$ from 
the CDF experiment~\cite{dms}, we obtain:
\begin{equation}
\label{aslg}
\Delta \Gamma_s \cdot \tan \phi_s = A_{SL}^s \cdot \Delta M_s = 
0.02 \pm 0.16~\mbox{ps}^{-1}.
\end{equation}

\begin{figure}[h!tb]
\begin{center}\includegraphics[%
  width=7.8cm,
  keepaspectratio,
  trim=0 40 0 0
  ]{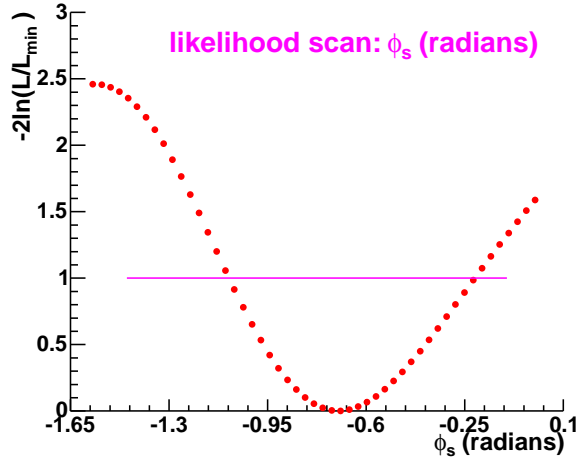}\end{center}
\caption{The likelihood scan versus $\phi_s$ for the
constrained fit (see text).}
\label{fig:bs_phiscan}
\end{figure}

We have repeated the fit 
to the \bsdec\ data, including the constraints from Eq.~\ref{aslg},
and from the world-average measurement of
$\tau_{\text {fs}}$ discussed earlier.
To illustrate the fit results and the impact of the constraints,
in Figs.~\ref{fig:dgvstau} -- \ref{fig:bs_contdgvsdphi_aslcombtausl}
we present  likelihood contours in three planes,
 $\Delta \Gamma_s$ versus $\overline {\tau}_s$,
 $\overline {\tau}_s$ versus  $\phi_s$, and
  $\Delta \Gamma_s$   versus  $\phi_s$, respectively.
The contours indicate error ellipses, $\Delta\ln({\cal L}) = 0.5$,
corresponding to the confidence level of 39\%.
The 4-fold ambiguity remains unresolved.
The likelihood profile as a function
of $\phi_s$ for the first solution listed in Eq.~\ref{solutions}
is shown in Fig.~\ref{fig:bs_phiscan}.
The extracted value of $\phi_s$ deviates from zero by 1.2 standard deviations.

In summary, for the solution with 
 $\phi_s <0$, $\cos \delta_1>0$ and $\cos \delta_2<0$,
we find the decay width difference and 
the CP-violating phase in the 
$B_s^0$ system to be:
\begin{eqnarray}  
\Delta \Gamma_s & = & 0.13 \pm 0.09~\rm {ps} ^{-1}, \nonumber \\
\phi_s & = & -0.70 ^{+0.47}_{-0.39}.
\end{eqnarray}
The measurement uncertainty is dominated by the limited statistics. 
The systematic uncertainties
include a variation of the background model in the analysis of the
decay \bsdec,
detector acceptance, 
and sensitivity
to the details of the track and vertex reconstruction. 
The results are consistent with the SM predictions~\cite{LN2006}.

\medskip
\input acknowledgement_paragraph_r2.tex   


\end{document}

%% file: list_of_authors_r2.tex
%
\author{                                                                      
V.M.~Abazov,$^{35}$                                                           
B.~Abbott,$^{75}$                                                             
M.~Abolins,$^{65}$                                                            
B.S.~Acharya,$^{28}$                                                          
M.~Adams,$^{51}$                                                              
T.~Adams,$^{49}$                                                              
E.~Aguilo,$^{5}$                                                              
S.H.~Ahn,$^{30}$                                                              
M.~Ahsan,$^{59}$                                                              
G.D.~Alexeev,$^{35}$
% ...
\\                                                                            
\vskip 0.30cm                                                                 
\centerline{(D\O\ Collaboration)}                                             
\vskip 0.30cm                                                                 
}                                                                             
\affiliation{                                                                 
\centerline{$^{1}$Universidad de Buenos Aires, Buenos Aires, Argentina}       
\centerline{$^{2}$LAFEX, Centro Brasileiro de Pesquisas F{\'\i}sicas,         
                  Rio de Janeiro, Brazil}                                     
\centerline{$^{3}$Universidade do Estado do Rio de Janeiro,                   
                  Rio de Janeiro, Brazil}                                     
\centerline{$^{4}$Instituto de F\'{\i}sica Te\'orica, Universidade            
                  Estadual Paulista, S\~ao Paulo, Brazil}                     
\centerline{$^{5}$University of Alberta, Edmonton, Alberta, Canada,           
                  Simon Fraser University, Burnaby, British Columbia, Canada,}
\centerline{York University, Toronto, Ontario, Canada, and                    
                  McGill University, Montreal, Quebec, Canada}                
\centerline{$^{6}$University of Science and Technology of China, Hefei,       
                  People's Republic of China}                                 
\centerline{$^{7}$Universidad de los Andes, Bogot\'{a}, Colombia}             
\centerline{$^{8}$Center for Particle Physics, Charles University,            
                  Prague, Czech Republic}                                     
\centerline{$^{9}$Czech Technical University, Prague, Czech Republic}         
\centerline{$^{10}$Center for Particle Physics, Institute of Physics,         
                   Academy of Sciences of the Czech Republic,                 
                   Prague, Czech Republic}                                    
\centerline{$^{11}$Universidad San Francisco de Quito, Quito, Ecuador}        
\centerline{$^{12}$Laboratoire de Physique Corpusculaire, IN2P3-CNRS,         
                   Universit\'e Blaise Pascal, Clermont-Ferrand, France}      
\centerline{$^{13}$Laboratoire de Physique Subatomique et de Cosmologie,      
                   IN2P3-CNRS, Universite de Grenoble 1, Grenoble, France}    
\centerline{$^{14}$CPPM, IN2P3-CNRS, Universit\'e de la M\'editerran\'ee,     
                   Marseille, France}                                         
\centerline{$^{15}$Laboratoire de l'Acc\'el\'erateur Lin\'eaire,              
                   IN2P3-CNRS et Universit\'e Paris-Sud, Orsay, France}       
\centerline{$^{16}$LPNHE, IN2P3-CNRS, Universit\'es Paris VI and VII,         
                   Paris, France}                                             
\centerline{$^{17}$DAPNIA/Service de Physique des Particules, CEA, Saclay,    
                   France}                                                    
\centerline{$^{18}$IPHC, IN2P3-CNRS, Universit\'e Louis Pasteur, Strasbourg,  
                   France, and Universit\'e de Haute Alsace,                  
                   Mulhouse, France}                                          
\centerline{$^{19}$IPNL, Universit\'e Lyon 1, CNRS/IN2P3, Villeurbanne, France
                   and Universit\'e de Lyon, Lyon, France}                    
\centerline{$^{20}$III. Physikalisches Institut A, RWTH Aachen,               
                   Aachen, Germany}                                           
\centerline{$^{21}$Physikalisches Institut, Universit{\"a}t Bonn,             
                   Bonn, Germany}                                             
\centerline{$^{22}$Physikalisches Institut, Universit{\"a}t Freiburg,         
                   Freiburg, Germany}                                         
\centerline{$^{23}$Institut f{\"u}r Physik, Universit{\"a}t Mainz,            
                   Mainz, Germany}                                            
\centerline{$^{24}$Ludwig-Maximilians-Universit{\"a}t M{\"u}nchen,            
                   M{\"u}nchen, Germany}                                      
\centerline{$^{25}$Fachbereich Physik, University of Wuppertal,               
                   Wuppertal, Germany}                                        
\centerline{$^{26}$Panjab University, Chandigarh, India}                      
\centerline{$^{27}$Delhi University, Delhi, India}                            
\centerline{$^{28}$Tata Institute of Fundamental Research, Mumbai, India}     
\centerline{$^{29}$University College Dublin, Dublin, Ireland}                
\centerline{$^{30}$Korea Detector Laboratory, Korea University,               
                   Seoul, Korea}                                              
\centerline{$^{31}$SungKyunKwan University, Suwon, Korea}                     
\centerline{$^{32}$CINVESTAV, Mexico City, Mexico}                            
\centerline{$^{33}$FOM-Institute NIKHEF and University of                     
                   Amsterdam/NIKHEF, Amsterdam, The Netherlands}              
\centerline{$^{34}$Radboud University Nijmegen/NIKHEF, Nijmegen, The          
                  Netherlands}                                                
\centerline{$^{35}$Joint Institute for Nuclear Research, Dubna, Russia}       
\centerline{$^{36}$Institute for Theoretical and Experimental Physics,        
                   Moscow, Russia}                                            
\centerline{$^{37}$Moscow State University, Moscow, Russia}                   
\centerline{$^{38}$Institute for High Energy Physics, Protvino, Russia}       
\centerline{$^{39}$Petersburg Nuclear Physics Institute,                      
                   St. Petersburg, Russia}                                    
\centerline{$^{40}$Lund University, Lund, Sweden, Royal Institute of          
                   Technology and Stockholm University, Stockholm,            
                   Sweden, and}                                               
\centerline{Uppsala University, Uppsala, Sweden}                              
\centerline{$^{41}$Physik Institut der Universit{\"a}t Z{\"u}rich,            
                   Z{\"u}rich, Switzerland}                                   
\centerline{$^{42}$Lancaster University, Lancaster, United Kingdom}           
\centerline{$^{43}$Imperial College, London, United Kingdom}                  
\centerline{$^{44}$University of Manchester, Manchester, United Kingdom}      
\centerline{$^{45}$University of Arizona, Tucson, Arizona 85721, USA}         
\centerline{$^{46}$Lawrence Berkeley National Laboratory and University of    
                   California, Berkeley, California 94720, USA}               
\centerline{$^{47}$California State University, Fresno, California 93740, USA}
\centerline{$^{48}$University of California, Riverside, California 92521, USA}
\centerline{$^{49}$Florida State University, Tallahassee, Florida 32306, USA} 
\centerline{$^{50}$Fermi National Accelerator Laboratory,                     
            Batavia, Illinois 60510, USA}                                     
\centerline{$^{51}$University of Illinois at Chicago,                         
            Chicago, Illinois 60607, USA}                                     
\centerline{$^{52}$Northern Illinois University, DeKalb, Illinois 60115, USA} 
\centerline{$^{53}$Northwestern University, Evanston, Illinois 60208, USA}    
\centerline{$^{54}$Indiana University, Bloomington, Indiana 47405, USA}       
\centerline{$^{55}$University of Notre Dame, Notre Dame, Indiana 46556, USA}  
\centerline{$^{56}$Purdue University Calumet, Hammond, Indiana 46323, USA}    
\centerline{$^{57}$Iowa State University, Ames, Iowa 50011, USA}              
\centerline{$^{58}$University of Kansas, Lawrence, Kansas 66045, USA}         
\centerline{$^{59}$Kansas State University, Manhattan, Kansas 66506, USA}     
\centerline{$^{60}$Louisiana Tech University, Ruston, Louisiana 71272, USA}   
\centerline{$^{61}$University of Maryland, College Park, Maryland 20742, USA} 
\centerline{$^{62}$Boston University, Boston, Massachusetts 02215, USA}       
\centerline{$^{63}$Northeastern University, Boston, Massachusetts 02115, USA} 
\centerline{$^{64}$University of Michigan, Ann Arbor, Michigan 48109, USA}    
\centerline{$^{65}$Michigan State University,                                 
            East Lansing, Michigan 48824, USA}                                
\centerline{$^{66}$University of Mississippi,                                 
            University, Mississippi 38677, USA}                               
\centerline{$^{67}$University of Nebraska, Lincoln, Nebraska 68588, USA}      
\centerline{$^{68}$Princeton University, Princeton, New Jersey 08544, USA}    
\centerline{$^{69}$State University of New York, Buffalo, New York 14260, USA}
\centerline{$^{70}$Columbia University, New York, New York 10027, USA}        
\centerline{$^{71}$University of Rochester, Rochester, New York 14627, USA}   
\centerline{$^{72}$State University of New York,                              
            Stony Brook, New York 11794, USA}                                 
\centerline{$^{73}$Brookhaven National Laboratory, Upton, New York 11973, USA}
\centerline{$^{74}$Langston University, Langston, Oklahoma 73050, USA}        
\centerline{$^{75}$University of Oklahoma, Norman, Oklahoma 73019, USA}       
\centerline{$^{76}$Oklahoma State University, Stillwater, Oklahoma 74078, USA}
\centerline{$^{77}$Brown University, Providence, Rhode Island 02912, USA}     
\centerline{$^{78}$University of Texas, Arlington, Texas 76019, USA}          
\centerline{$^{79}$Southern Methodist University, Dallas, Texas 75275, USA}   
\centerline{$^{80}$Rice University, Houston, Texas 77005, USA}                
\centerline{$^{81}$University of Virginia, Charlottesville,                   
            Virginia 22901, USA}                                              
\centerline{$^{82}$University of Washington, Seattle, Washington 98195, USA}  
}                                                                             

%% file: acknowledgement_paragraph_r2.tex
%
We thank the staffs at Fermilab and collaborating institutions, 
and acknowledge support from the 
DOE and NSF (USA);
CEA and CNRS/IN2P3 (France);
FASI, Rosatom and RFBR (Russia);
CAPES, CNPq, FAPERJ, FAPESP and FUNDUNESP (Brazil);
DAE and DST (India);
Colciencias (Colombia);
CONACyT (Mexico);
KRF and KOSEF (Korea);
CONICET and UBACyT (Argentina);
FOM (The Netherlands);
PPARC (United Kingdom);
MSMT (Czech Republic);
CRC Program, CFI, NSERC and WestGrid Project (Canada);
BMBF and DFG (Germany);
SFI (Ireland);
The Swedish Research Council (Sweden);
Research Corporation;
Alexander von Humboldt Foundation;
and the Marie Curie Program.